\documentclass[runningheads]{llncs}
\usepackage[T1]{fontenc}
\usepackage{graphicx}
\usepackage{amsmath}
\begin{document}
\title{Towards Audience-Driven Real-Time Data Storytelling}
%
%
\author{Angelica Lo Duca\orcidID{0000-0002-5252-6966}}
\authorrunning{Lo Duca}
%
\institute{IIT-CNR, Pisa 56124, Italy
\email{angelica.loduca@iit.cnr.it}\\
}
\maketitle              
\begin{abstract}
This paper introduces the concept of Audience-Driven Real-Time Data Storytelling (ADRDS), a framework that utilizes participatory and social sensing to enable audiences to actively shape the presentation of data stories. At its core, ADRDS employs a pre-authored decision tree, where the presenter defines the data story through a branching three-act arc (setup, conflict, and resolution). Each node contains a question, branching rules, thresholds for coverage, consensus, and confidence intervals, as well as one or more potential calls to action (CTAs). Audience members connect to the ADRDS framework via a URL on their mobile devices and answer questions in real time. ADRDS evaluates responses against the decision tree in real-time, following strong branches when thresholds are satisfied or activating fallback paths when signals are inconclusive. ADRDS then selects the final action consistent with the traversed path and the audience’s demonstrated readiness, subject to fairness safeguards.
ADRDS transforms presentations of data stories into audience-shaped decision processes, where live data actively drives the narrative. This work paves the way towards a new form of data storytelling, where the final data story depends on the audience's background and interests.

\end{abstract}
\section{Introduction}
Data storytelling conveys the results of a data analysis process through a narrative, combining data and rhetorical structure to make insights interpretable and memorable~\cite{Dykes2019}. Traditionally, such stories are authored offline for a predefined or curated audience: the storyteller prepares visuals and text in advance, tailoring content according to assumptions about the audience’s knowledge and motivations~\cite{Segel2010}. 
In practice, audiences vary in their objectives, interests, and dispositions~\cite{LoDuca2025}, suggesting that storytellers should adapt narratives not only to the data but also to the audience present in real time. Prior work on game live streaming shows that audiences can actively influence live experiences~\cite{Striner2021}. Systems capable of sensing feedback through polls, sensors, or behavioral cues could therefore infer audience preferences and dynamically adjust both narrative structure and presentation style.
Recent work has also used augmented reality to enrich data storytelling. InSituTale~\cite{Takahira2025} embeds narratives in physical objects, while TangibleNet~\cite{Takahira2025b} supports real-time manipulation of network visualizations through tangible interactions. However, these approaches primarily enhance presenter expressiveness rather than audience agency.

Mobile phones and laptops have further enabled computer-augmented performer–audience interaction, often termed live participation~\cite{Nelimarkka2016,Nelimarkka2018}, through mechanisms such as live tweeting and backchannel chats. However, such technologies require careful orchestration to create meaningful experiences~\cite{Reeves2005} and avoid distracting from the performance~\cite{Cerratto-Pargman2014}. Although HCI research has increasingly examined collocated social interaction~\cite{salovaara2021augmenting}, and interactive narrative has produced intelligent systems approaches~\cite{Riedl2013}, prior work has not explored audiences as active contributors to data storytelling. This gap motivates adaptive frameworks that integrate audience participation into narrative progression.

This paper proposes Audience-Driven Real-Time Data Storytelling (ADRDS), a framework that integrates live polling and participatory social sensing as key components of a data story progression. Rather than treating polls as isolated engagement tools, ADRDS uses them as branching signals in a narrative structure. Presenters define the main data story in advance. Then, they refine the story for different potential audiences, thus generating as many refined stories as possible for each audience. Finally, they use a Decision Control Engine (DCE) to organize the scenes of the stories into a decision tree. During the data delivery process, participants respond via mobile devices; these responses are aggregated and fed to DCE, which evaluates the distributions of responses against branching rules, consensus thresholds, and fairness constraints and chooses the next narrative beat. The presenter decides the next scene of the story based on the output generated by the DCE. However, they retain override control to preserve coherence.
This paper introduces the ADRDS framework and presents a lightweight prototype implementation together with a simulation-based evaluation. The results show the potential effectiveness of ADRDS in enabling real-time adaptive data storytelling driven by audience feedback, highlighting its potential to foster more engaging and responsive presentations.

\section{Related Work}
Audience response systems, adaptive presentation technologies, and participatory sensing have progressively changed the role of audiences in live events, turning them from passive observers into active participants. Building on these developments, the ADRDS framework takes a further step by allowing collective audience input to shape narrative decisions directly as the story unfolds.

Audience Response Systems (ARS), such as Poll Everywhere~\cite{sarvary2017}, Socrative~\cite{Pryke2020}, Kahoot!~\cite{Kohnke2022}, and Mentimeter~\cite{Little2016}, are widely used in classrooms and public events to foster participation and engagement. Prior work shows that live polls can improve student attention and create a more interactive environment~\cite{price2022}. Similarly, Salovaara \textit{et al.} highlight how mobile-based voting and backchannel tools are increasingly adopted in professional event productions~\cite{salovaara2021augmenting}. However, in both cases, audience input is limited to generating real-time statistics or stimulating discussion, while the presentation itself follows a predetermined storyline. 

Recent work in adaptive slideware has explored how presentations can dynamically adjust to situational context. \textit{Elastica} adapts slide animations to the presenter’s voice and gestures~\cite{Cao2024}. \textit{Presentation A11y} provides real-time feedback to help presenters make their slides accessible for visually impaired audiences~\cite{Peng2021}. 
Edge \textit{et al.} propose global editing techniques for collaboratively restructuring presentations, maintaining visual coherence while accommodating content changes~\cite{Edge2015}.
SketchStory~\cite{Lee2013} demonstrated how freeform sketching can be leveraged to create more engaging data stories, extending the author’s expressive control but without incorporating audience-driven adaptation.

Participatory and physiological sensing has been studied to understand audience engagement in live performances. Han \textit{et al.} analyzed electrodermal activity (EDA) and heart-rate variability (HRV) signals of dance audiences to link physiological arousal with narrative moments~\cite{Han2023}. Similarly, Kosmyna and Maes introduced \textit{AttentivU}, a wearable EEG-based system that provides real-time biofeedback to help listeners maintain attention during lectures~\cite{Kosmyna2019}. These systems demonstrate the potential of passive audience sensing to capture engagement and, in some cases, to support individual self-regulation. 

Prior work has demonstrated the value of audience response systems for engagement, adaptive presentation systems for tailoring delivery, and participatory sensing for monitoring audience reactions.
ADRDS extends this line of work by integrating collective audience responses directly into the narrative logic of the presentation. In the proposed framework, passive feedback is transformed into active control. Rather than merely informing or supporting the presenter, audience participation becomes the central mechanism that determines the progression of the story and its eventual outcomes. This positions ADRDS as an evolution of adaptive slideware, from systems that adapt presentation content or style to systems where the narrative itself is co-authored in real-time by the audience.

\section{Audience-Driven Real-Time Data Storytelling}

Data storytelling combines data and narrative to build compelling stories that invite the audience to action \cite{Dykes2019}. Different types of data stories can be built, depending on the audience and the objective \cite{Ojo2018}. However, in all the cases, a well-designed data story should have a plot and some characters \cite{Dasu2024}. Similar to a classical novel or movie, a data story can be organized following the three-act structure~\cite{LoDuca2025}, the hero's journey~\cite{Wei2025}, the Freytag pyramid~\cite{Yang2022}, the DIKW pyramid~\cite{LoDuca2025b}, or other narrative models. 

Traditionally, a data story is designed offline: analysts or communication specialists select relevant datasets, build visualizations, and craft narratives aimed at a specific target audience. This process is usually prepared in advance, assuming a static audience profile that is known or projected a priori. Once delivered, the story remains fixed and does not adapt to changes in audience composition or reactions.

\subsection{Types of Audiences}

The effectiveness of data storytelling depends strongly on the intended audience~\cite{Dykes2019}. Prior research and practice often distinguish between at least three major types: (1) the \textit{general public}, who require accessibility, clarity, and emphasis on key takeaways; (2) \textit{decision makers}, who demand concise insights, actionable implications, and strong recommendations; and (3) \textit{technical staff}, who expect methodological transparency, detailed data, and reproducibility~\cite{LoDuca2025}.
Beyond these functional categories, audiences can also be classified by attitude or disposition~\cite{Cingi2023}. For instance, some audiences may be \textit{hostile} (skeptical or resistant), \textit{interested} (engaged and receptive), or \textit{neutral}. Each classification influences the tone and level of detail and the persuasive strategies that the storyteller may adopt.
Traditional approaches to data storytelling assume that the presenter knows their audience in advance, so they prepare an a priori data story targeted to that specific audience.

\subsection{Real-Time Data Storytelling}

This paper extends the concept of data storytelling into the domain of \textit{real-time} audience interaction. The paper introduces \textit{Audience-Driven Real-Time Data Storytelling (ADRDS)}, a system in which the data story is dynamically shaped during a live presentation according to feedback provided by the audience. Unlike traditional approaches that rely on static, predefined narratives, ADRDS treats audience input as a continuous stream of participatory sensing signals. These signals are collected during the session and used to guide the trajectory of the presentation.

Conceptually, real-time data storytelling could be entirely emergent, with the narrative constructed on the fly during the live presentation. In practice, however, presenters benefit from defining a set of possible scenarios, narrative branches, and calls to action in advance. This preparation ensures that while the story remains flexible and responsive to audience feedback, the presenter retains coherence, control, and the ability to steer the session toward meaningful outcomes.

\section{Architecture}
\begin{figure}[t]
    \centering
    \includegraphics[width=0.9\columnwidth]{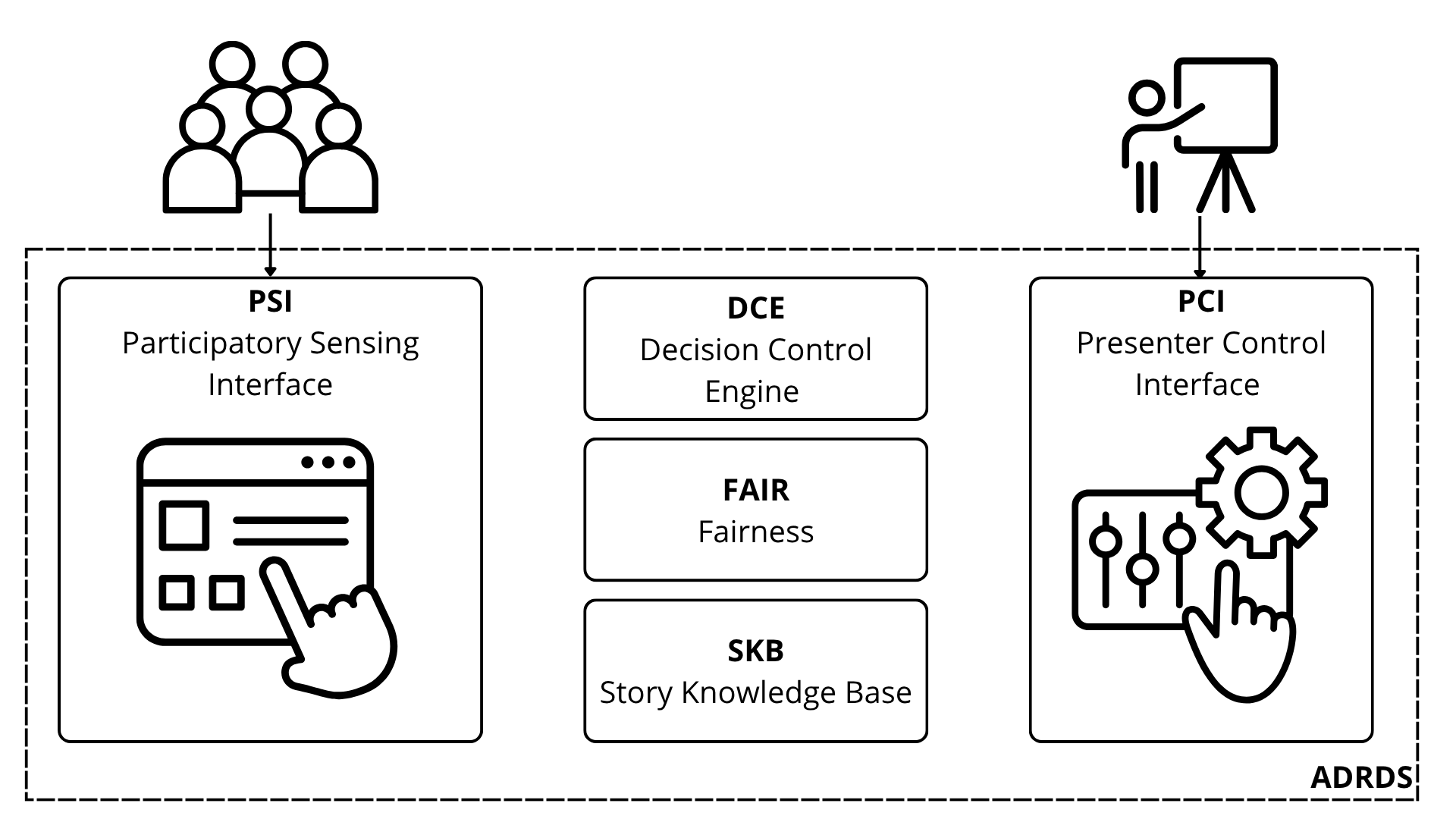}
    \caption{The ADRDS architecture}
    \label{fig:architecture}
\end{figure}

The ADRDS architecture integrates participatory sensing with adaptive narrative generation to enable audience-driven storytelling in real-time presentation environments. The framework consists of five core modules (Figure \ref{fig:architecture}): (i) the \textit{Participatory Sensing Interface (PSI)}, (ii) the \textit{Decision Control Engine (DCE)}, (iii) the \textit{Story Knowledge Base (SKB)}, (iv) the \textit{Presenter Control Interface (PCI)}, and (v) the \textit{Fairness} (FAIR). 

\paragraph{System Model.}
The ADRDS system operates in a distributed sensing environment comprising two principal entities: the presenter and the audience. The presenter $P$ defines the narrative scaffold and facilitates the session, while the audience $A = \{a_1, a_2, \ldots, a_n\}$ provides participatory sensing signals via the PSI. The DCE processes collective responses and communicates decisions to the PCI, through which the presenter validates progression. Narrative content is retrieved from the SKB, which stores all possible paths and variants of the story. 

\paragraph{Participatory Sensing Interface (PSI).}

The PSI enables audience members to connect to the system via mobile devices, ensuring easy access through URLs, QR codes, short links, or other forms of connection. It presents questions authored by the presenter and collects responses in real time. Collected signals are transmitted to the DCE in real time, forming the input for branching decisions.

\paragraph{Decision Control Engine (DCE).}

The DCE implements the story’s decision tree. 
The story structure is modeled as a directed tree $T = (N, E)$ where $N$ represents decision nodes and $E$ represents narrative transitions. Each node $n_i \in N$ encapsulates a question $q_i$ with response options $O_i = \{o_1, o_2, \ldots, o_k\}$, a branching function $f: O_i \to N$ that maps response outcomes to subsequent nodes, and threshold parameters $(\tau_c, \tau_{cons}, \tau_{ci})$ governing transition conditions. Terminal nodes contain candidate calls to action $CTA_i = \{cta_1, cta_2, \ldots, cta_m\}$ selected based on the traversed narrative path and audience readiness indicators.

For each question $q_i$, the system collects responses $R_i = \{r_1, r_2, \ldots, r_n\}$ where each $r_j \in O_i$. The response distribution is characterized by proportions:
\begin{equation}
p_k = \frac{|\{r_j : r_j = o_k\}|}{|R_i|}
\end{equation}
where $p_k$ represents the proportion of the audience selecting option $o_k$.

Coverage measures participation adequacy as:
\begin{equation}
\text{Coverage} = \frac{|R_i|}{|A|}
\end{equation}
and must exceed the \textit{coverage threshold} $\tau_c$, i.e., 
the minimum fraction of the audience required to provide a 
valid response at a decision node.

Consensus identifies the dominant response through:
\begin{equation}
p_{\max} = \max\{p_1, p_2, \ldots, p_k\}
\end{equation}
and must satisfy the \textit{consensus threshold} $\tau_{cons}$, 
i.e., the minimum proportion of responses supporting the 
majority option.
These metrics ensure that narrative decisions are based on sufficient audience participation and clear preference signals.

To quantify uncertainty in audience signals, ADRDS employs bootstrap confidence intervals. For the winning proportion $p_{\max}$, the 95\% confidence interval is computed through bootstrap resampling as:
\begin{equation}
CI_{0.05} = [p^*_{0.025}, p^*_{0.975}]
\end{equation}
where $p^*_\alpha$ represents the $\alpha$-quantile of the bootstrap distribution. The confidence width is defined as:
\begin{equation}
w_{ci} = p^*_{0.975} - p^*_{0.025}
\end{equation}
providing a measure of statistical certainty in the audience response.
The \textit{confidence interval threshold} $\tau_{ci}$ specifies 
the maximum allowable width, ensuring sufficient statistical 
certainty in the audience signal.

\paragraph{Threshold-Based Decision Logic.}
A narrative branch transition occurs when all threshold conditions are satisfied:
\begin{equation}
\text{Branch}(n_i \to n_j) \Leftrightarrow \text{Coverage} \geq \tau_c \land p_{\max} \geq \tau_{cons} \land w_{ci} \leq \tau_{ci}
\end{equation}
This formalization ensures that narrative progression only occurs when sufficient audience participation (coverage), clear preference (consensus), and statistical confidence (narrow confidence interval) are achieved simultaneously.
If thresholds are not met, the DCE triggers fallback procedures: neutral story branches or micro-questions to gather additional evidence. The resulting decision is sent to the PCI for validation.
At present, fallback procedures are not formally defined; their specification and evaluation are left for future work.

\paragraph{Presenter Control Interface (PCI).}

The PCI serves as the human-in-the-loop checkpoint. The presenter visualizes the DCE’s proposed transition, with context on participation, consensus, and confidence. The presenter can approve or override the suggested branch. Once validated, the PCI queries the SKB to fetch the next narrative element consistent with the decision.

\paragraph{Story Knowledge Base (SKB).}

The SKB stores the complete set of pre-authored narrative components. Stories can be implemented using classical three-act arcs or alternative dramaturgical models, as long as all possible branches are explicitly encoded. This paper focuses on the three-act structure of a data story. However, the described procedure can also be extended to other narrative models. The three-act structure assumes that a data story is organized in three acts: (I) setup, where the background or context information is presented; (II) conflict, where the main insight in the data is offered to the audience; and (III) resolution, where the solution and a possible CTA are given. 
To accommodate diverse audiences, the SKB must contain multiple intertwined variants. For example, if the target includes both technical and executive audiences, at least two parallel versions of the story must be implemented, with branching points allowing adaptive selection during the session.
Ideally, the SKB should also contain the slides to be presented for each scene of the story, ensuring the system is completely synchronized with the live presentation.

\paragraph{Fairness.}

At the beginning of a session, all participants are considered \emph{unknown}. After an initial profiling question, the audience is progressively segmented into categories (e.g., technical, executive, general). This segmentation allows the system to monitor whether all parts of the audience are adequately represented in decision-making.

Let $S$ be the set of audience segments, with estimated size $\widehat N_s$ and observed responders $R_s$. The coverage of segment $s$ is
\[
c_s = \frac{R_s}{\widehat N_s}.
\]
Fairness is measured through a \emph{Coverage Equity Score (CES)}, which quantifies how evenly participation is distributed:
\[
CES = 1 - \frac{1}{|S|}\sum_{s \in S}\big|c_s - \overline{c}\big|,
\quad
\overline{c}=\frac{1}{|S|}\sum_{s \in S} c_s.
\]
A branch decision is considered fair when (i) each segment reaches a minimum coverage threshold $c_s \ge \tau_c^{seg}$, (ii) the CES exceeds an equity threshold $\tau_{ces}$, and (iii) the fraction of participants still in the unknown bucket $N_U/N$ remains below $\tau_u$ or the worst–case equity bound $CES_{\min}$ (assigning all unknowns to the weakest segment) is still acceptable.

If these conditions are not satisfied, the system refrains from branching and instead triggers fallback strategies such as neutral story paths, micro–questions to engage underrepresented segments, or presenter mediation.

\section{Prototype Implementation}
To evaluate the feasibility of ADRDS, a lightweight prototype was developed. The system was implemented in Python as open-source code\footnote{https://github.com/alod83/ADRDS}, using a modular architecture designed for rapid experimentation and extensibility.

\begin{figure}[t]
    \centering
    \includegraphics[width=0.6\columnwidth]{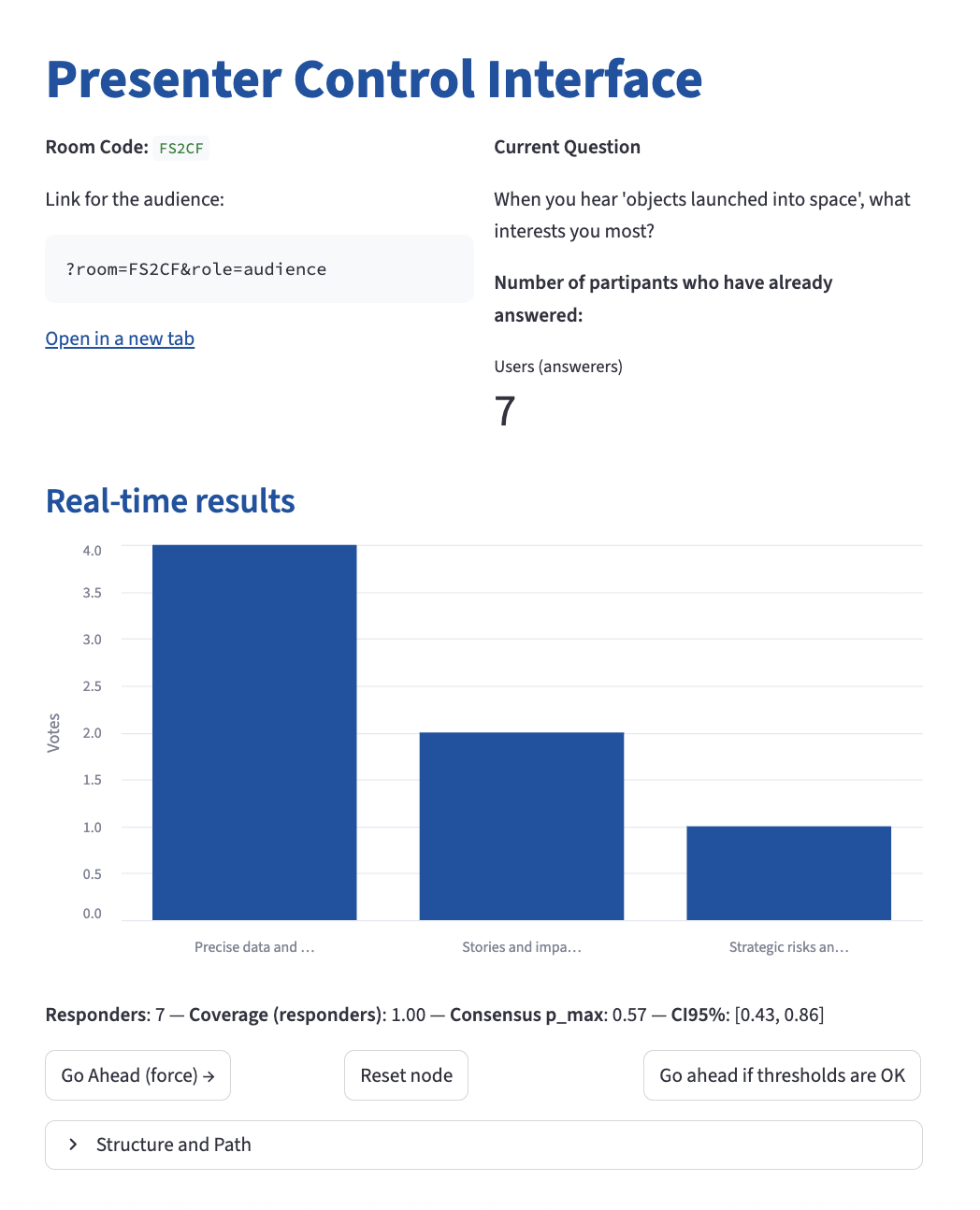}
    \caption{The Interactive Interface for the presenter}
    \label{fig:ui-presenter}
\end{figure}

\subsection{DCE and SKE}

The DCE manages the dynamic progression of the story and is implemented as a backend Python script. The SKE is encoded in an external JSON file (\texttt{story.json}), which defines nodes, branching conditions, and terminal calls to action (CTAs). Each node specifies the narrative text, the available options presented to the audience, and the mapping of options to subsequent nodes. This representation allows non-technical users (e.g., presenters) to design branching storylines without modifying source code. 
The following code snippet shows the structure of a node in the JSON file (populated with sample values):
\begin{small}
\begin{verbatim}
{"start": "setup",
  "nodes": {
    "setup": {
      "id": "setup",
      "text": "Question1",
      "options": [
        {"id": "technical", "text": "Option1"},
        {"id": "general", "text": "Option2"},
        {"id": "executive", "text": "Option3"}
      ],
      "next_map": {
        "technical": "tech_act1",
        "general": "gen_act1",
        "executive": "exec_act1"}},
    "tech_act1": {...},
    "gen_act1": {...},
    "exec_act1": {...}}
\end{verbatim}
\end{small}

Ideally, the SKE should also define for each node the current scene of the data story, such as the location of the physical associated slide in the file system. However, at the moment, this feature has not been implemented yet.
The DCE loads the JSON structure at runtime, maintains the current state of the session, and applies threshold-based logic to evaluate audience responses.

\subsection{PCI and PSI}

The PCI and PSI are implemented through a single web interface that can operate in two complementary modes: \textit{presenter mode} and \textit{audience mode}.

In presenter mode (PCI), the presenter is provided with a dashboard that displays the current state of the active question, including the number of participants who have submitted responses and the distribution of votes (Figure \ref{fig:ui-presenter}). The presenter can choose to proceed to the following narrative step once a predefined threshold has been met or force the transition even if the threshold has not been reached. This control mechanism ensures that the presentation remains coherent and progresses at an appropriate pace. Additionally, the presenter can create and manage new sessions, allowing different presentations or storylines to be run independently.

\begin{figure}[t]
    \centering
    \includegraphics[width=0.60\columnwidth]{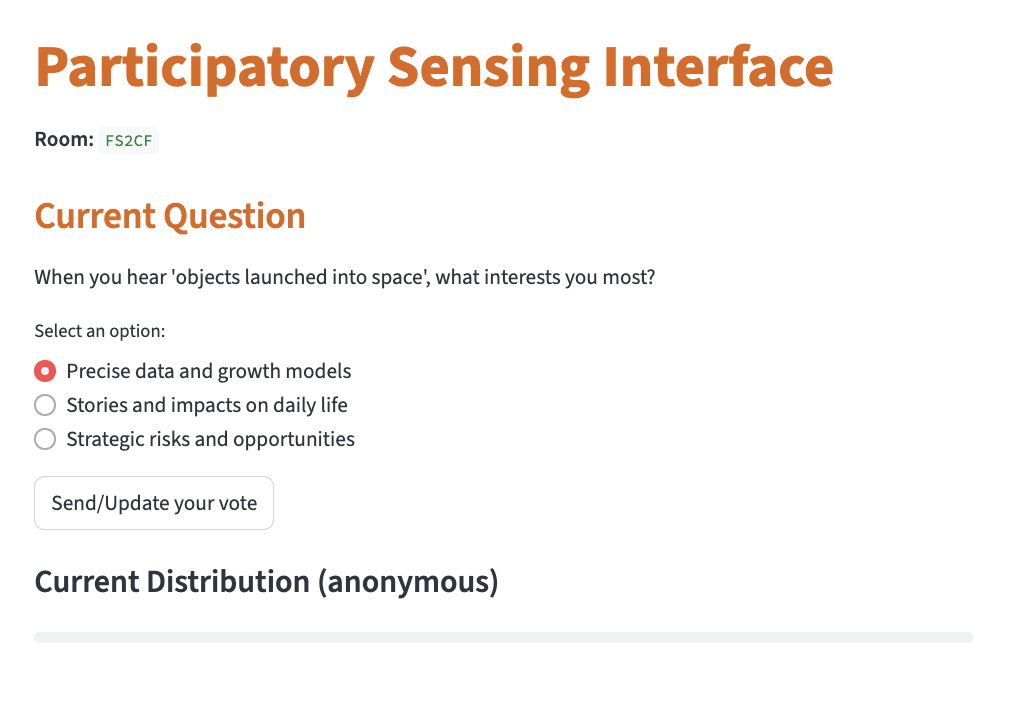}
    \caption{The Interactive Interface for the audience}
    \label{fig:ui-audience}
\end{figure}

In audience mode (PSI), participants are shown the active question as it becomes available (Figure \ref{fig:ui-audience}). Each participant can submit a response using a simple interface optimized for mobile devices. Responses can also be updated within the active time window, ensuring flexibility for participants who wish to reconsider their input. The system automatically synchronizes the audience interface with the presenter’s dashboard, guaranteeing that all participants interact with the same current question and that responses are immediately aggregated for processing by the narrative engine.

\subsection{Extensibility}

Although the current prototype uses multiple-choice questions as the primary interaction modality, the architecture is extensible. Future extensions may incorporate free-text input, gesture recognition, or physiological sensing data. Similarly, additional branching logic can be defined to incorporate audience sentiment, temporal dynamics, or fairness constraints, enabling richer adaptation of the narrative.

This prototype demonstrates the core feasibility of ADRDS: encoding branching narratives, collecting real-time audience input, and producing dynamically evolving stories that conclude with role-specific CTAs.

\section{Simulated Scenario}
To assess the feasibility of ADRDS, this paper describes a simulated case study with a realistic data story and a synthetic evaluation of the system scalability and fairness.

\subsection{The Data Story}
\begin{figure}[t]
    \centering
    \includegraphics[width=0.6\columnwidth]{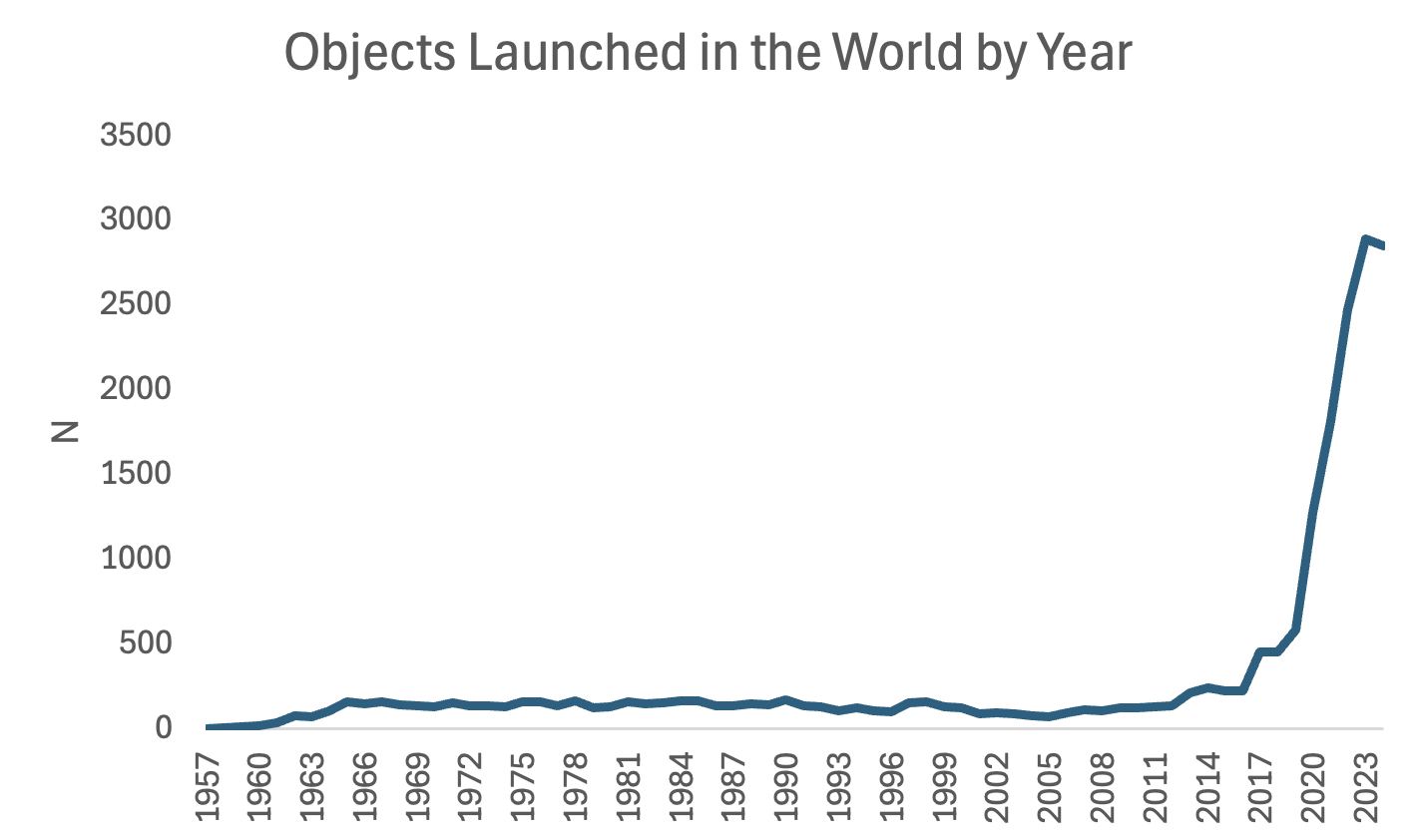}
    \caption{Objects Launched in the World by Year}
    \label{fig:act1}
\end{figure}

\begin{figure}[t]
    \centering
    \includegraphics[width=0.6\columnwidth]{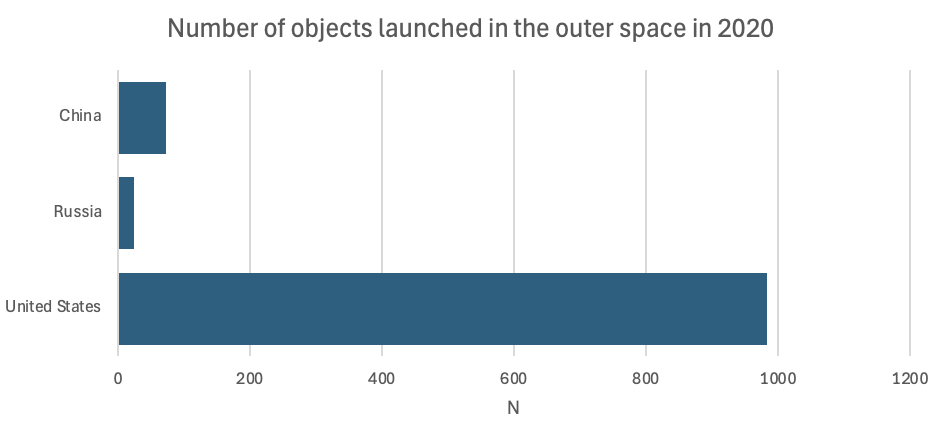}
    \caption{Number of objects launched into outer space in 2020}
    \label{fig:act3}
\end{figure}

The scenario focuses on the evolution of space launches, using data from \cite{owid-space-exploration-satellites}. The narrative is organized into three acts: (I) the number of objects launched into outer space until 2019 (see the trend line in Figure \ref{fig:act1} up to 2019), (II) the sharp increase in launches starting in 2020 (see the trend line Figure \ref{fig:act1} since 2020), and (III) the identification of the main driver of this increase, namely the surge of launches from the United States (see Figure \ref{fig:act3}). The complete JSON file of the story is distributed with the ADRDS open-source prototype$^1$.

Three parallel versions of this story are authored: one for a \textit{technical audience}, another for the  \textit{general audience}, and one for an \textit{executive audience}. For simplicity, each act is represented as a single decision node with only two branching options.

During the simulation, the presenter uses a set of slides that are dynamically updated based on the output of the DCE. The implementation of the slides is out of the scope of this paper. Ideally, the slides should be embedded in the SKB.

The simulation begins with an entry node (\texttt{setup}), which probes the audience with the question: \textit{``When you hear 'objects launched into space', what interests you most?''}. The available options (``Precise data and growth models'',  ``Stories and impacts on daily life'' and ``Strategic risks and opportunities'') act as a branching point. Based on aggregated responses, the DCE selects the subsequent path: either a technical branch (\texttt{tech\_act1}) a general branch (\texttt{gen\_act1}), or an executive branch (\texttt{exec\_act1}).

For participants with a strategic focus, the storyline highlights risks, opportunities, and decision-making. The opening node (\texttt{exec\_act1}) contrasts \textit{risk management} with \textit{opportunity capture}. Follow-up branches (\texttt{exec\_act2\_risk} and \texttt{exec\_act2\_opportunity}) address either regulatory uncertainty, supply chain fragility, and geopolitical competition, or new markets, partnerships, and efficiency gains. Each path concludes with an executive-level call-to-action (CTA) prompting reflection on resilience, governance, and competitive positioning. This branch centers on how leaders can convert space-related disruption into strategy.

For participants oriented toward quantitative insights, the narrative progresses through successive analytical questions. For instance, \texttt{tech\_act1} asks whether the launch rate up to 2019 appeared stable, while \texttt{tech\_act2\_stable} and \texttt{tech\_act2\_growing} explore whether the 2020 spike represents an anomaly or a new trend. Each branch embeds a call-to-action (CTA) encouraging reflection on forecasting, anomaly detection, or trend monitoring. 

For participants preferring narrative impact, the general branch (\texttt{gen\_act1}) asks whether space was filling slowly or quickly before 2019. Subsequent nodes, such as \texttt{gen\_act2\_slow} and \texttt{gen\_act2\_fast}, discuss whether the 2020 boom is isolated or marks a new era. The third act focuses on identifying the main drivers of the new space race, with options leading to CTAs highlighting public support, private entrepreneurship, or collaborations as transformative forces.


At each decision node, the PSI collects responses and the DCE applies coverage, consensus, and confidence thresholds. The proposed branch is displayed in the PCI, where the presenter validates or overrides the choice. The SKB then provides the narrative element corresponding to the validated branch. The process continues until a terminal node is reached, where a CTA is issued. The prototype highlights how the framework allows a single presentation to bifurcate into audience-specific narratives while ensuring structural coherence.

\subsection{Scalability and Fairness}

To assess the scalability and fairness of ADRDS, this paper presents a synthetic simulation of audience responses across varying audience sizes and decision tree depths. The simulation models how individual participants decide whether to respond and which option (A vs. B) they select at each decision node.  

\paragraph{Audience Composition.}
We tested two scenarios of audience segmentation. In the \emph{balanced} scenario, the synthetic audience is partitioned into three archetypal segments (technical (35\%), executive (25\%), and general public (40\%)) 
In the \emph{unbalanced} scenario, the distribution was skewed towards the general public, with proportions technical (10\%), executive (20\%), and general (70\%), and base preferences of 0.62, 0.55, and 0.50, respectively. In both scenarios, the base preferences for option A were set to 0.62, 0.55, and 0.50, respectively. This setting was designed to capture heterogeneity of preferences and participation in a relatively symmetric way.  

The exact proportions are not intended as empirical ground truth, but rather as modeling choices to enable fairness and scalability analysis under different audience profiles. Different distributions can be easily tested within our simulation framework. Each participant is assigned to a segment at the start of the simulation, and the total number of participants is adjusted to exactly match the desired audience size $N$.

At each decision node, each participant independently decides whether to respond with probability:
\[
p_\text{resp}(N) = \max(0.5,\,0.85 - 0.0002 \cdot N).
\]
This captures the intuition that participation decreases slightly as the audience grows larger. The effective \emph{coverage} of a node is then computed as the fraction of the audience who submitted a response.  

Participants who respond then vote between options A and B. To introduce variability across nodes, we add a small Gaussian drift $\delta \sim \mathcal{N}(0,0.05)$, clipped between $0.05$ and $0.95$, to the base probability. Thus, for segment $s$ at node $i$, the probability of voting A is:
\[
p_{A}^{(s,i)} = \text{clip}\big(\text{base\_pref}[s] + \delta_i,\;0.05,\,0.95\big).
\]
Votes are then sampled as $V_A^{(s)} \sim \text{Binom}(R_s, p_{A}^{(s,i)})$, where $R_s$ is the number of respondents from segment $s$. The proportion of A votes is $p_A = V_A / \sum_s R_s$, with $p_B = 1-p_A$.  

At each node, the DCE evaluates whether to branch based on three thresholds. 
We experimented with varying values of both the coverage threshold $\tau_c$ 
and the consensus threshold $\tau_{cons}$, starting from 0.40 and increasing 
in steps of 0.1. The analysis showed that the best trade-off between 
branch success rate and fairness was obtained when both thresholds were set 
to 0.50. Therefore, in the remainder of our experiments, we fixed 
$\tau_c = \tau_{cons} = 0.50$. We also set $\tau_{ci} = 0.5$.

\paragraph{Branch Success Rate}
Scalability and fairness are measured through the branch success rate (BSR). A node is considered a \emph{successful branch} if all thresholds are satisfied. The BSR is defined as the fraction of nodes in a simulated tree where branching occurs successfully. This metric reflects how reliably ADRDS can sustain interactive storytelling under different audience sizes $N$ and tree depths $d$. We tested different values of $N$, ranging from 20 to 2000, and different values of $d$, ranging from 2 to 10. 

\begin{figure}[t]
    \centering
\includegraphics[width=0.8\columnwidth]{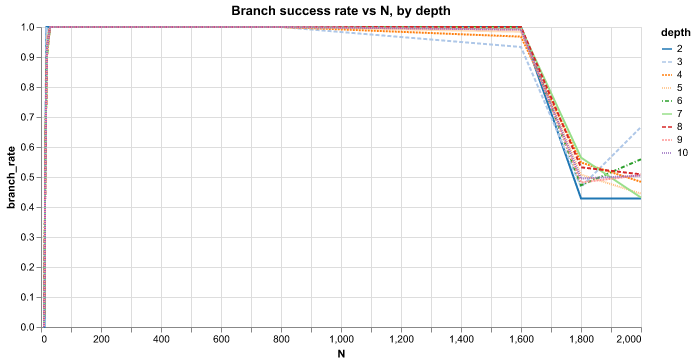}
    \caption{Balanced scenario: Branch success rate vs audience size.}
\label{fig:balanced_branch_vs_N}
\end{figure}

\begin{figure}[t]
    \centering
\includegraphics[width=0.8\columnwidth]{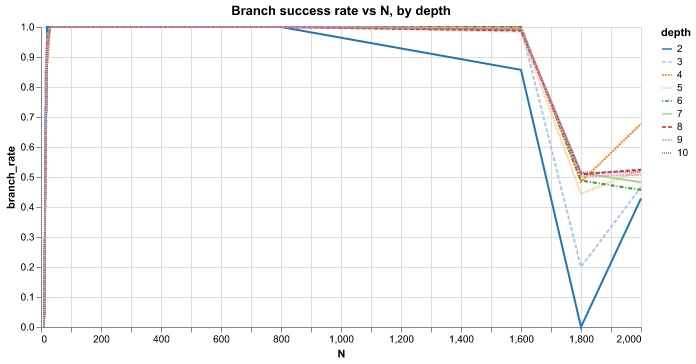}
    \caption{Unbalanced scenario: Branch success rate vs audience size.}
    \label{fig:unbalanced_branch_vs_N}
\end{figure}

\paragraph{Scalability}
Scalability is the ability of the system to maintain reliable performance and decision accuracy as the audience size and narrative tree depth increase. 
Initially, we tested scalability with the fairness module disabled.
Figures~\ref{fig:balanced_branch_vs_N} and~\ref{fig:unbalanced_branch_vs_N} summarize the results of our simulations under both the balanced and the unbalanced audience configurations. In both scenarios, we enforced a minimum audience size of 10 nodes; therefore, values below 10 are not represented, and the figures appear to drop to zero for very small $N$.  

Overall, the BSR is close to~1.0 for all audience sizes greater than 10, and this level of performance is maintained even for very large audiences up to 1600 participants. However, differences emerge between the two audience configurations at larger scales. In the unbalanced scenario, performance begins to degrade gradually from $N=800$ upwards, especially for shallow trees (depth 2--3), where BSR can drop below 0.4. In the balanced scenario, the system remains stable up to $N=1600$, but beyond this threshold a sharp degradation occurs: the BSR declines significantly, reaching values around 0.4 at the largest tested size.  

\begin{figure}[t]
    \centering
\includegraphics[width=0.8\columnwidth]{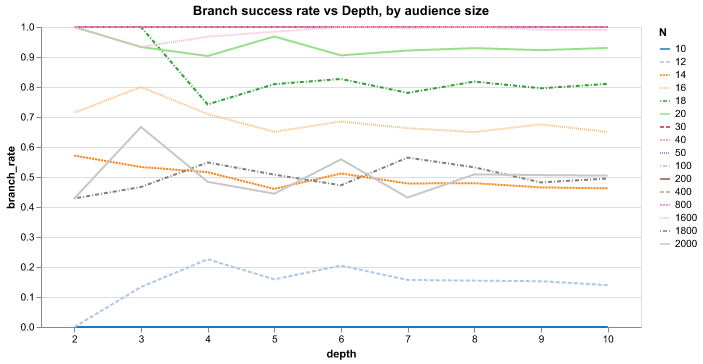}
    \caption{Balanced Scenario: Branch success rate vs depth.}
\label{fig:balancedBranchDepth}
\end{figure}

\begin{figure}[t]
    \centering
\includegraphics[width=0.8\columnwidth]{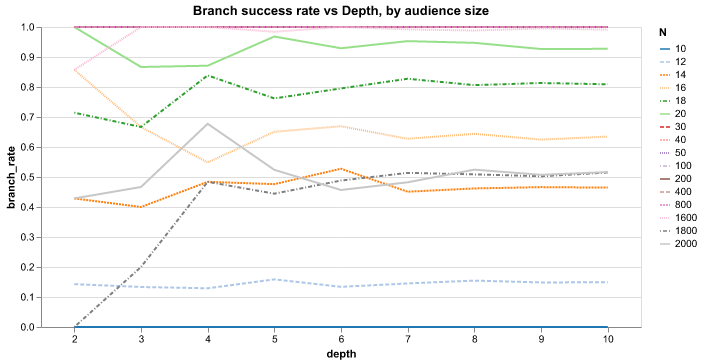}
    \caption{Unbalanced Scenario: Branch success rate vs depth.}
    \label{fig:unbalancedBranchDepth}
\end{figure}

Figures~\ref{fig:balancedBranchDepth} and~\ref{fig:unbalancedBranchDepth} report the BSR as a function of tree depth for the balanced and unbalanced audience scenarios, respectively.  
In the balanced case, the highest BSR is obtained for an audience size of $N=800$. More generally, for depths greater than 4, the BSR remains relatively stable and shows only minor variations as the tree becomes deeper. This indicates that once a sufficient number of branching opportunities has been reached, additional depth does not significantly affect the system’s ability to sustain reliable decisions.  
In the unbalanced case, the best performance is achieved for audiences below $N=800$, with the BSR consistently higher in this range. Beyond that point, performance declines, although the degradation is moderate. As in the balanced case, for tree depths larger than 4 the BSR stabilizes and remains largely independent of the depth parameter.  

Overall, these results highlight that the ADRDS system can sustain branching reliably across a wide range of depths. Audience composition influences the optimal size at which maximum BSR is reached: balanced audiences peak at larger sizes, whereas unbalanced audiences favor smaller populations.

\begin{figure}[t]
    \centering
\includegraphics[width=0.8\columnwidth]{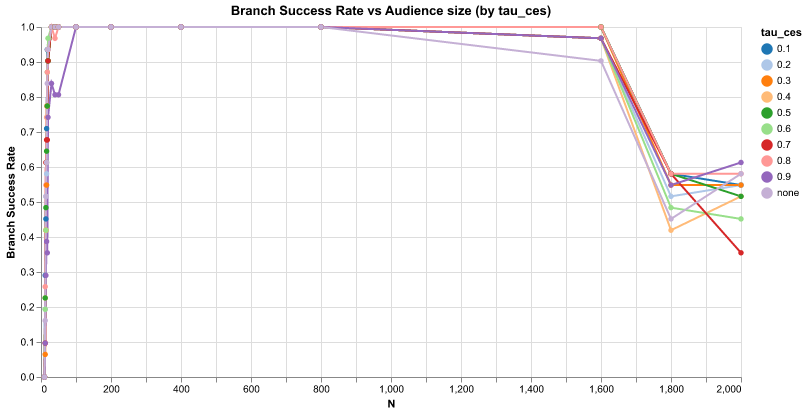}
    \caption{Balanced Scenario: Branch success rate vs Audience size (by $tau_{ces}$.}
\label{fig:balancedBranchTauCes}
\end{figure}

\begin{figure}[t]
    \centering
\includegraphics[width=0.8\columnwidth]{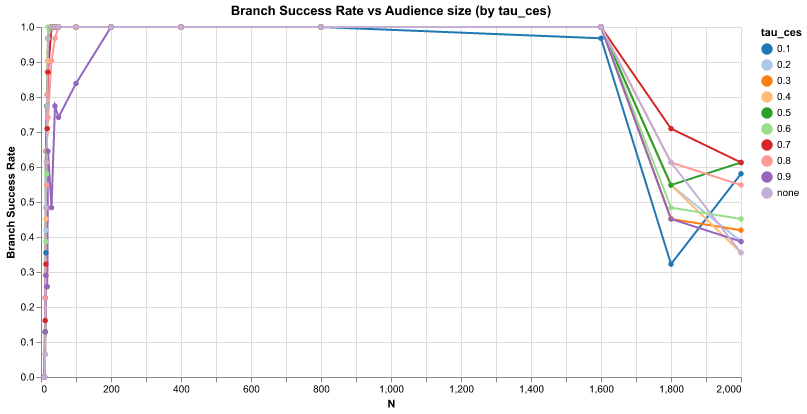}
    \caption{Unbalanced Scenario: Branch success rate vs Audience size (by $tau_{ces}$.}
    \label{fig:unbalancedBranchTauCes}
\end{figure}

\paragraph{Fairness}
To test fairness, we measured how the branch success rate (BSR) varied with
audience size when varying the equity threshold $\tau_{ces}$.
Figures~\ref{fig:unbalancedBranchTauCes} and~\ref{fig:balancedBranchTauCes}
report the BSR for the unbalanced and balanced scenarios,
respectively.  

As discussed earlier, BSR is irrelevant for $N<10$ and naturally degrades at very large scales, even when fairness is disabled. 
This behavior reflects intrinsic scalability limits of ADRDS, not the 
fairness module itself. Our fairness evaluation therefore focused on whether 
stricter $\tau_{ces}$ values would exacerbate this degradation.  

In the \emph{balanced} scenario, BSR quickly reaches values close to~1.0 for 
$N \geq 30$ and remains robust up to $N=800$ and beyond. Varying $\tau_{ces}$ 
has only a marginal effect in this case: even with higher fairness thresholds, 
performance remains stable, indicating that a symmetrically composed audience 
is naturally resilient to fairness constraints. Some degradation is visible 
only at extreme scales ($N \geq 1600$), where BSR drops to around 0.5.  

In the \emph{unbalanced} scenario, the overall trend is similar, but stricter 
$\tau_{ces}$ values accelerate the degradation observed beyond $N=800$. 
In particular, with high fairness thresholds, BSR can decline to 0.4 or below, 
especially when shallow trees are considered. This suggests that when one 
segment dominates the audience, ensuring equity across groups becomes 
increasingly difficult at scale.  

Overall, these results show that fairness thresholds introduce only a 
moderate additional cost. The main source of degradation lies in scalability 
itself, which occurs regardless of fairness. Nonetheless, fairness constraints 
remain valuable to ensure inclusivity, especially in balanced populations 
where their impact on performance is minimal.

\section{Discussion}
The proposed ADRDS framework demonstrates promising scalability properties. 
Our simulations indicate that the system sustains reliable branching decisions 
up to audiences of approximately 800 participants. Importantly, this is already 
well beyond the typical size of live presentations, which usually involve only 
tens or at most a few hundred attendees. Even at scales of up to 1600 participants, 
the system remains functional, although with gradually decreasing performance. 
This suggests that ADRDS can be realistically deployed in real-world scenarios 
without significant risk of failure due to audience size. 
Another important finding is that enabling the fairness module does not 
significantly degrade performance. The system continues to sustain branching decisions effectively even when fairness constraints are enforced across heterogeneous audience segments. This property ensures that ADRDS can scale while maintaining inclusivity and representativeness.

\subsection{Challenges}
While the current prototype demonstrates the scalability of the DCE, other scalability concerns warrant careful consideration. 
Ensuring fair representation across diverse audience segments presents ongoing challenges. Digital divide issues may systematically exclude participants without modern mobile devices or sufficient technical literacy, creating biased samples that misrepresent the broader audience perspective. Although the current system describes some mechanisms for detecting or correcting such participation biases, further research is required.

Cultural and linguistic barriers also affect participation equity. The prototype assumes shared language proficiency and cultural context for interpreting questions, potentially disadvantaging non-native speakers or individuals from different cultural backgrounds. Real-time translation or culturally adaptive questioning mechanisms represent important areas for future development.
Another important challenge concerns security and ethics. From a technical perspective, the system must guarantee audience anonymity while preventing malicious behaviors such as repeated voting or identity spoofing, which could manipulate results and force the presenter to follow unintended narrative paths. Ensuring robust authentication and one-vote-per-user mechanisms is therefore essential to preserve trust. Beyond technical safeguards, ADRDS also raises ethical concerns: by adapting stories to audience preferences, there is a risk of reinforcing biases or merely telling people what they want to hear, rather than fostering critical reflection~\cite{Sangalang2013}. 
This suggests that ADRDS must balance responsiveness to audience input with safeguards that preserve narrative coherence and critical reflection.

\subsection{Limitations}
Despite these encouraging results, several limitations remain. 
First, the evaluation relies on synthetic tests rather than user studies with 
real participants. While simulations provide scalability insights, empirical 
validation with actual audiences is essential to assess engagement, usability, 
and overall experience. Second, the synchronization between ADRDS and 
presentation slides was considered out of scope in this work. In the current 
prototype, slides must be managed externally, which introduces a risk of 
misalignment between the evolving story and the presented content. Ideally, 
slides should be directly integrated into the ADRDS system and retrieved 
from the SKB, ensuring coherence and seamless presentation. 
In addition, at present, fairness is treated as
balancing participation across segments. This does not necessarily guarantee fair influence over the story,
nor does it protect minority perspectives when the system selects majority-supported directions.
Therefore, future work should explore alternative fairness objectives, such as
minority-preserving branching or weighted representation.
Long-term effects of audience-driven storytelling also remain unexplored. While the framework aims to increase engagement and persuasive impact, it may inadvertently reduce critical thinking by emphasizing consensus over individual reflection. The psychological and social implications of participatory narrative construction warrant systematic investigation.

\section{Conclusions and Future Work}
ADRDS represents a shift from static, pre-authored presentations to 
interactive, audience-driven storytelling. By integrating participatory 
sensing and structured narrative control, the system enables real-time 
adaptation of stories based on audience input. The results described in this paper suggest that 
ADRDS is feasible and scalable to audiences significantly 
larger than typical presentation sizes. This paradigm fosters a sense of collective 
ownership of the narrative and opens new opportunities for engagement, 
personalization, and inclusivity in live events. 

Future work will extend ADRDS in several key directions. 
First, privacy-preserving mechanisms must be incorporated to ensure that 
audience contributions remain anonymous and protected, for instance, through 
aggregation, anonymization, or noise injection. Second, integration of the 
presentation layer with the SKB will be pursued, allowing 
slides and other media to be dynamically loaded as narrative elements. Third, empirical 
evaluation with real audiences will be essential to complement the synthetic 
simulations, validating engagement, usability, and robustness in diverse 
venues and network conditions. Fourth, ADRDS could be compared with existing polls mechanisms. Finally, ADRDS could be enhanced with 
artificial intelligence, for example, by integrating large language models 
to generate adaptive story variations, enrich branching logic, and support 
dynamic, context-sensitive calls to action, while preserving narrative 
coherence and ethical safeguards.
%
%
%
\bibliographystyle{splncs04}
\bibliography{sections/refs}

\end{document}